# Do "bad" citations have "good" effects?

Honglin Bao[1]* and Misha Teplitskiy[2]*

[1] Harvard Business School, [2] School of Information, University of Michigan, Ann Arbor

* To whom correspondence may be addressed:

Misha Teplitskiy, tepl@umich.edu, 105 S State St, Ann Arbor, MI 48109, USA

Honglin Bao, hbao@hbs.edu, 50 North Harvard Street, Allston, MA 02163, USA

# Abstract

The scientific community discourages authors of research papers from citing papers that did not influence them. Such "rhetorical" citations are assumed to degrade the literature and incentives for good work. While a world where authors cite only substantively appears attractive, we argue that mandating substantive citing may have underappreciated consequences on the allocation of attention and dynamism in scientific literatures. We develop a novel agent-based model in which agents cite substantively and rhetorically. Agents first select papers to read based on their expected quality, read them and observe their actual quality, become influenced by those that are sufficiently good, and substantively cite them. Next, agents fill any remaining slots in the reference lists by (rhetorically) citing papers that support their narrative, regardless of whether they were actually influential. By turning rhetorical citing on-and-off, we find that rhetorical citing increases the correlation between quality and citations, increases citation churn, and reduces citation inequality. This occurs because rhetorical citing redistributes some citations from a stable set of elite-quality papers to a more dynamic set with high-to-moderate quality and high rhetorical value. Increasing the size of reference lists, often seen as an undesirable trend, amplifies the effects. In sum, rhetorical citing helps deconcentrate attention and makes it easier to displace incumbent ideas, so whether it is indeed undesirable depends on the metrics used to judge desirability.

# 1. Introduction

Citations are widely used in science to measure the impact of papers and researchers. The assumption underlying this evaluative use of citations is that scientists generally follow the norm that citations acknowledge intellectual debts to prior work (Baldi, 1998; Zuckerman, 1987). If most citations are indeed of this "substantive" type, the enormously laborious and subjective task of assessing the impact of research becomes a relatively straightforward and objective one – just count the citations. However, decades of research show that many citations reflect motivations that have little, if anything, to do with acknowledging intellectual debts (Bornmann & Daniel, 2008). According to the social constructivist theory of citing, authors cite works to persuade readers and reviewers, regardless of whether the works *influenced* the authors (Gilbert, 1977; Cozzens, 1989; Latour & Woolgar, 2013; Nicolaisen, 2007). We refer to these as "rhetorical" citations (Teplitskiy *et al.*, 2022). Rhetorical citations help authors persuade audiences and navigate the publishing process in several ways. Some rhetorical citations provide context (Allen, 1997) or differentiate the citer's contributions from prior works by criticizing them (Catalini *et al.*, 2015). Some rhetorical citations are "coerced" during peer review (Wilhite & Fong, 2012). Consistent with these relatively superficial uses of the literature, studies find that in many cases authors misrepresent the claims of the papers they cite (Cobb *et al.*, 2023) and do not read them thoroughly (Simkin & Roychowdhury, 2005). Authors can usually differentiate whether a citation was made to acknowledge intellectual debts or for other purposes (Teplitskiy *et al.*, 2022), supporting the distinction between substantive and rhetorical citations.

Authors need not be indifferent to a paper's quality when considering citing it rhetorically. For example, prominent papers may bolster the citer's claims more than obscure papers (Rubin & Rubin, 2021), a phenomenon sometimes called "persuasion by naming-dropping" (White, 2004; Frandsen & Nicolaisen, 2017). Nevertheless, quality plays at most an indirect role in rhetorical citing – the author's own perception of a paper's quality is much less relevant than how potential readers perceive its quality, because the higher the latter, the more persuasive the paper.

Rhetorical citing is generally seen as an undesirable practice that corrupts the literature and incentives for future research (Penders, 2018). The view is so common and uncontroversial that the journal *Nature Genetics* has gone as far as to explicitly warn that manuscripts citing rhetorically will be rejected (*Nature Genetics* Editorials, 2017). Even aside from official policies, some argue that rhetorical citing is a signal of low-quality work (Petric, 2007; Rose, 2014).

The view that rhetorical citing is undesirable implicitly compares the current world with rhetorical citing to a counterfactual world without it. Yet a rigorous comparison between the two worlds does not exist and, as we argue below, is worth conducting because which world is preferable is unclear *ex ante*. This is because the counterfactual world with only substantive citing is likely to be one in which attention is concentrated on only the few best papers, and their advantage becomes locked-in over time. Thus, the first contribution of this paper is to conduct such a comparison. To do so, we use a novel, behavioral model of the citing process, which is the paper's second contribution. The model combines substantive and rhetorical motivations to cite, along with cognitively realistic search and reading practices. Formulating a comprehensive theory of citing has been a challenge in information and library science for many decades (Cronin, 1981). The main existing theories – normative and social constructivist – have been criticized as incomplete as stand-alone theories (Cozzens, 1989; Nicolaisen, 2007), and there is little consensus on an overall synthetic theory (Bornmann & Daniel, 2008; Tahamtan & Bornmann, 2019). This paper proposes such a synthesis.

The guiding intuition is that researchers' cognitive constraints prevent them from reading all relevant papers. Instead, when seeking papers to read and possibly cite substantively, researchers will use heuristics like citation counts to identify papers they expect to have the highest quality. Their attention is then likely to concentrate on a small number of the most cited (or otherwise highest-status) works and only papers from this set would get cited (Chu & Evans, 2021; Parolo *et al.*, 2015). The feedback loop between citing and more citing would lock-in certain papers as winners and make it difficult for new ideas to receive attention. In contrast to substantive citing, when citing rhetorically researchers consider *other, non-quality* related factors, *i.e.* papers' rhetorical value, and this weighting of non-quality factors redistributes some attention away from the highest-status works and gives other papers a chance to be cited. In other words, rhetorical citing helps to weaken the feedback loop and makes science more dynamic, at least as measured in terms of citations.

Empirically, comparing scientific communities with and without rhetorical citing is challenging. First, there may not exist any communities without rhetorical citing (Teplitskiy *et al.* 2022). Our model addresses this challenge by simulating artificial communities with arbitrary levels of rhetorical citing. This enables us to turn rhetorical citing "on" and "off" and measure its effects. Second, even if such communities did exist, classifying citations as substantive or rhetorical is difficult. One approach uses machine learning and training data from third-party labelers (Jurgens *et al.*, 2018; Nicholson *et al.*, 2021), but how well such labels correspond to authors' actual motivations is unclear. Teplitsky *et al.* (Teplitsky *et al.*, 2022) use a survey approach, asking authors why they cited a specific paper. However, surveys may have response biases, and are difficult to do at large scale. We address this challenge by building into our model two citation types - substantive and rhetorical. Third, to establish whether particular citing practices have deleterious effects on the health of a scientific community it is important to identify compelling measures of health. After surveying the literature we identified three outcomes and the associated measures that are often seen as capturing aspects of health: citation-quality correlation, citation churn (*i.e.*, the replacement of reference list and disruption of the established canon), and citation inequality.

The rest of this paper is structured as follows. Section 2 reviews empirical stylized facts about scientific information search, reading, and citing practices. Section 3 describes three metrics commonly taken to reflect the health of scientific communities. Section 4 describes the agent-based model. Section 5 presents the results of the main comparison, as well as the effects of changing the reading and citing budgets and literature sizes. Sections 6 and 7 discuss the findings and conclusions.

## 2. Search, reading and citing practices

We first describe several stylized facts about how researchers search for, read, and cite papers that underlie the model developed in Section 4. These stylized facts, some of which are rather cynical, are not ones we endorse or seek to normalize. They are simply practices that, to the best of our knowledge, are well supported by empirical studies.

### Reading

Researchers do not read all potentially relevant papers[1], but select which ones to read strategically (Tenopir *et al.*, 2015; Renear & Palmer, 2009). A key criterion for selecting among relevant papers is quality (or "fitness" (Wang *et al.*, 2013)) – scientists prioritize *reading* (even if not citing) the best papers (Wang & Domas White, 1999). Identifying the best papers to read is challenging since quality can be hard to discern at a glance. Researchers instead use heuristics (Lerman *et al.*, 2017; Petersen *et al.*, 2014). One commonly used heuristic is a paper's or author's status, which is assumed to proxy quality (Simcoe & Waguespack, 2010; Wang & Soergel, 1998; Azoulay *et al.*, 2013). Citation count is one component of status (Teplitskiy *et al*., 2022). Thus, when scientists search for papers to read, they are likely to prioritize the highly cited ones. Heuristic-based selection can cause attention and citations to increasingly concentrate among the highest cited works, due to the feedback loop between current selections and future selections (Peterson *et al.*, 2014).

---
[1] For conciseness we refer to various types of research works, *e.g.*, papers, books, as "papers."

## Citing substantively

Citing to acknowledge intellectual debts is often referred to as the "normative" theory of citing (Nicolaisen, 2007; Bornmann & Daniel, 2008). This theory posits that there is a norm in science to acknowledge intellectual debts (Merton, 1973) and researchers hold themselves to this norm. To use more colloquial terminology, we call normative citations "substantive." Substantive citing presupposes reading, since it is difficult to be substantively influenced by a work without knowing its contents. The styled facts around selection and cumulative advantage in reading described above should also apply to substantive citing. For example, if a researcher tends to select papers with high perceived quality to read, she will tend to only substantively cite (and be influenced by) papers with high perceived quality (not actual quality).

## Citing rhetorically

Authors also cite works to persuade readers, regardless of whether the works had substantive influences on them (Nicolaisen, 2007; Gilbert, 1977). We refer to this type of citing as "rhetorical." Because rhetorical citing does not require influence, it does not require close (and, at the extreme, any) reading. The non-necessity of reading is supported by several lines of evidence. Studies comparing what citers of papers claim those papers say *vs.* what they actually say show frequent disagreements and distortions (Horbach *et al.*, 2021; Leigh Star, 2010; Mizruchi & Fein, 1999), with 9.5% of a sample of psychology citations being outright mischaracterizations of the underlying papers (Cobb *et al.*, 2023). Cases of very specific mistakes in what a paper is taken to claim or in its actual reference string are difficult to explain except through a lack of careful reading (Rekdal, 2014; Simkin & Roychowdhury, 2005). Lastly, in surveys authors report citing papers without carefully reading them and/or without being influenced by them (Teplitskiy *et al.*, 2022).

Researchers are likely to select papers for rhetorical citations based on their "rhetorical value," which can be affected by time-invariant characteristics like quality (Gilbert, 1977) and publication outlet, but also time-varying characteristics like citations and the status of the authors (Azoulay *et al.*, 2013; Rubin & Rubin, 2021). A paper's rhetorical value is likely to increase the more it is cited (Lerman *et al.*, 2017). High citation counts make papers appear to be of higher quality, more significant, more novel, and more generalizable (Teplitskiy *et al.*, 2022; Hargens, 2000; Allen, 1997). The relative rhetorical value of a paper can thus increase (or decrease) over time.

Rhetorical value of a paper is also likely to differ from researcher-to-researcher, based on how well the paper's claims match the researcher's objectives. For example, when two researchers write on a controversial topic and cite rhetorically, a paper that supports one side of the controversy may be rhetorically valuable for one researcher and less valuable for the other (Shwed & Bearman, 2010). The rhetorical value of a paper is thus likely to vary between researchers, depend on some time-invariant characteristics like quality and fit, and change over time as it accrues status, citations, or becomes obsolete.

# 3. Metrics of community health

In order to rigorously compare scientific communities with and without rhetorical citing, we identified in the literature three dimensions of community "health" – citation-quality correlation, citation churn, and citation inequality. We do not take these dimensions as exhaustive (we return to this point in the *Discussion*), and only claim that they are commonly discussed. For each of the three metrics we develop a hypothesis, based on the intuition that in a world without rhetorical citing, attention and citations would be highly concentrated among the few elite-quality papers. In contrast, because rhetorical citing depends on factors beyond quality, it redistributes some attention and citations to good-but-not-elite papers. Low-quality papers are generally not cited substantively or rhetorically because they are not good or persuasive. We then consider the moderating effects on these characteristics of the reading budget (how many papers researchers read), citing budget (how many references they may put in a paper), and the literature size (the number of papers relevant to a scientist in a period of time).

## Metric 1: Citations-quality correlation

Despite long-standing critiques of citations (and metrics derived from them, like journal impact factor and *h*-index) as a measure of quality, in practice, they are often used as such by administrators and analysts (Bornmann & Daniel, 2008). Researchers themselves perceive more highly cited papers as of higher quality (Teplitskiy *et al.*, 2022). Regardless of whether citations should be used as a proxy for quality, *if* they are, it is arguably better that they are a good proxy rather than a bad one. Consequently, we assume that the higher the correlation between quality and citations, the better for the community.

Rhetorical citing can make the correlation stronger. With only substantive citing, the citation distribution becomes in effect bimodal, with high-quality papers receiving all of them and others receiving 0 citations. With rhetorical citing, researchers consider a variety of factors beyond quality and, consequently, citations are more evenly and proportionally distributed across low- to high-quality papers. The more proportional distribution will have a stronger correlation. **Hypothesis 1**: Rhetorical citations increase the citation-quality correlation.

## Metric 2: Citation churn

As knowledge in a field evolves, some ideas receive increasing support and, eventually, may become taken-for-granted or "blackboxed" (Latour, 1987). In a healthy community, if new ideas arise that are better than the old ones, they should be recognized as such, and the old ones should be displaced. Such disruptive ideas are associated with higher novelty (Lin *et al.*, 2022) and are predictive of the highest level of recognition in the scientific community, like the Nobel prize (Wu *et al.*, 2019). A robust amount of turnover in reference lists across time, which we call "citation churn," may thus indicate a healthy evolution of published knowledge. In contrast, if the same set of papers remains the highest cited decade after decade, the community may experience stagnation. Indeed, empirical work suggests that such stagnation is on the rise (Chu &

Evans, 2021; Evans, 2008; Park *et al.*, 2023; Parolo *et al.*, 2015). Rhetorical citing may increase citation churn by reducing lock-in. Different researchers may find different papers rhetorically useful, *e.g.*, supporting their own claims, and not concentrate their attention and citations on only elite-quality papers. The more equal distribution of citations then weakens the feedback loop from current to future citations.

**Hypothesis 2**: Rhetorical citing increases the citation churn.

## Metric 3: Citation inequality

Scholarship has long found that the distribution of citations is highly skewed (Price, 1963), and that inequality appears to be increasing with time (Nielsen & Andersen, 2021; Gomez *et al.*, 2022). This degree of inequality is often described as problematic (Nielsen & Andersen, 2021), and possibly indicative of stagnation (Chu & Evans, 2021; Parolo *et al.*, 2015). While the optimal degree of inequality is debated, there is evidence that the realized degree is affected by factors such as technology (Evans, 2008) and even seemingly irrelevant factors like choice architecture (Feenberg *et al.*, 2016). Here, we follow this latter literature in investigating the effect of rhetorical citing on citation inequality, without taking a strong position on what amount is optimal for science.

In a world with substantive citing only, citations would be concentrated among the highest-quality papers, and that concentration would increase via the feedback loop of researchers choosing highly cited works to read and citing them yet more. Rhetorical citing may decrease citation inequality because researchers select papers to cite rhetorically based on a variety of idiosyncratic factors like person-specific rhetorical value, not only quality.

**Hypothesis 3**: Rhetorical citing decreases citation inequality.

## Moderating effects of literature size, and reading and citing budgets

The relationships between rhetorical citing and the metrics above may be moderated by several characteristics of a scientific community. First, consider literature size, or the number of papers relevant to a particular researcher. While literature size can change dramatically over time and topic, scientists' cognitive constraints are relatively stable. The stability of cognition implies that the number of papers scientists are capable of reading and being influenced by is also relatively stable. The larger the literature size, the smaller the fraction of papers a scientist will read and cite. Consequently, the larger the literature (while keeping citing budgets and other factors fixed), the more unequal the citation distribution. Relatedly, the more unequal the citation distribution, the lower the quality-citations correlation.

**Hypothesis 4**: the larger the literature size, the greater the citation inequality and the lower the quality-citation correlation.

We also expect the metrics of community health to be affected by researchers' reading and citing "budgets" – the number of papers they can realistically read, and are expected by specific fields and outlets to cite, respectively. While researchers' reading practices are difficult to measure at scale, the typical number of references in a paper is easily observable, and varies widely across fields and time (Andersen, 2023). We expect researchers to know their reading and citing budgets. Citing budgets are of particular interest from a policy perspective, as they can be and are routinely set by publishing outlets[2]. The more a researcher reads – the higher the reading "budget" – the more likely she is to give attention to non-elite-quality papers. However, if the researcher is limited to a small set of references, the slots will continue to go to papers that have had a substantive influence, presumably those of elite quality, and the additional papers will not be cited. When the reference list is expanded, for example by allowing rhetorical citations, researchers have more opportunities to populate it with non-elite-quality but rhetorically useful papers. Larger citing budgets, but not reading budgets alone, should thus reduce citation inequality, improve the correlation between citations and quality, and increase citation churn.

**Hypothesis 5A**: Increasing the citing budget (length of reference list) increases citation-quality correlation and citation churn, and reduces citation inequality.

**Hypothesis 5B:** The independent effect of increasing the reading budget is minor.

## 4. Formalizing the model

This section takes the search, reading and citing practices described in Section 2 and formalizes them into a family of models that vary from the simplest, *e.g.*, homogeneous agents, to more complex and realistic, *e.g.*, heterogeneous agents. When agents are homogeneous, they perceive the quality and rhetorical values of papers identically, and when they are heterogeneous, a paper may have a higher topical or rhetorical "fit" and therefore higher value to one agent than another. A key feature of adoption models like these is the strength of the "reinforcement" or "social influence" process – the degree to which an agent's adoption in a time period is determined by adoption by other agents in a previous period (Denrell & Liu, 2021). We parametrize the strength of reinforcement with two parameters $\alpha$ and $\beta$, for substantive and rhetorical values respectively. Additionally, modelers sometimes include mechanisms that prevent runaway reinforcement, which can lead to the unrealistic phenomenon of guaranteed adoption (van de Rijt, 2019; Zuckerman, 2012). For example, some assume that the reinforcement effect decays over time (Wang *et al.*, 2013). To simplify the model, we do not impose a ceiling on the strength of reinforcement, which is clearly unrealistic in the long-run. Consequently, a key scope condition of our models is that they model citing in the short-to-medium run.

---
[2] For example, *Nature* recommends that authors of research articles limit themselves to 50 references. https://www.nature.com/nature/for-authors/formatting-guide. Accessed 2023-04-02.

Our models follow the logic of classic threshold adoption models (Granovetter, 1978; Centola, 2011), where agents choose a work to cite if it exceeds some person-specific threshold, with one crucial change: we allow for multiple *types* of adoption. The way agents, papers, and agent-paper are modeled is enumerated below, and Table 1 summarizes the parameters and their distributions.

**Table 1.** Parameters used in the model.

| Level of analysis | Parameter name | Symbol | Ranges of values | Notes |
|---|---|---|---|---|
| Paper $i$ | Underlying quality | $q_i$ | [0,1] | Distribution: *Beta*(1,6)<br>Robust to other distributions, see Appendix 1.1: value distributions. |
| Agent $j$ | Threshold | $\tau_j$ | [0,1] | Distribution: *Uniform*(0,1).<br>Helps determine how many references in the agent's reference list are substantive *vs.* rhetorical.<br>Robust to the normal distribution, see Appendix 1.2: threshold. For homogenous citers, $\tau_j = 0.5$ |
| Paper $i$ - agent $j$ | Perception error | $\varepsilon_{i,j}$ | ≈[-.15,.15] | Distribution: *Normal*(0, 0.05). Min and max values are appx. ±3*SD = ±0.15.<br>Robust to other distributions, see Appendix 1.3: perception error. |
| | Fit | $fit_{i,j}$ | [-.1,.1] | Distribution: *Uniform*(-.1,.1).<br>Robust to more/less variant fits and their effects are much lower than rhetorical citing, see Appendix 1.4: fit. Person-specific for heterogeneous citers, identical for homogeneous citers. |
| | Perceived quality | $s_{i,j,t}$ | [0,2] | $s_{i,j,t} = q_i + fit_{i,j} + \varepsilon_{i,j} + \alpha \times c_{i,t}$<br>where $\alpha \times c_{i,t}$ captures the effect of citation count on perceived quality. The maximum citation premium $\alpha \times c_{i,t} = 1$ (see reinforcement strength). Values of $q_i + fit_{i,j} + \varepsilon_{i,j}$ that are >1 or <0 are set to either 0 or 1, respectively. |
| | Overall rhetorical value | $r_{i,j,t}$ | [0, 1.6] | $r_{i,j,t}\|unread = r_{i,j} + \beta * s_{i,j,t}$<br>$r_{i,j,t}\|read = r_{i,j} + \beta * (s_{i,j,t} - \varepsilon_{i,j})$<br>Composed of an underlying rhetorical value $r_{i,j}$ component and a component that depends on perceived quality. Distribution: *Beta*(1,6). Person-specific for heterogeneous citers, identical for homogeneous citers. Max value: $1+2*\beta = 1.6$.<br>Robust to other underlying distributions and other values of $\beta$, see Appendix 1.1: value distributions and Appendix 1.5: reinforcement strengths. |
| Reinforcement strength | Effect of citation count on perceived quality | $\alpha$ | .001 | Robust to other reinforcing strengths, see Appendix 1.5: reinforcement. The maximum citation premium is 1000 (max cites) * 0.001 = 1 |

| Level of analysis | Parameter name | Symbol | Ranges of values | Notes |
|---|---|---|---|---|
| | Effect of perceived quality on rhetorical value | β | .3 | Robust to other values of β, see Appendix 1.5: reinforcement. The maximum perceived quality premium is 2 (max perceived quality)*0.3 = 0.6. |
| Moderating effects | Literature size | $N$ | 200-800 | |
| | Reading budget | $m$ | 50-150 | |
| | Citing budget | $n$ | 20-100 | |
| Time steps | | $t$ | 1000 | |

## Characteristics of agent $j$

- **Quality threshold:** represented by $\tau_j$ for agent $j$. Agents substantively adopt (cite) a paper when its value to them exceeds a threshold. Homogeneous agents have identical thresholds, while heterogeneous agents differ in their thresholds. For agents with very high thresholds, very few papers will meet that bar for a substantive citation, so more of the reference list will be composed of rhetorical citations. In this way, thresholds help determine the composition of the reference list, even when its overall length is fixed.
- **Reading budget $m$:** Agents can read a maximum of $m$ papers, which they select based on perceived (not necessarily actual) quality.
- **Citing budget $n$:** Agents can cite a maximum of $n$ papers.

## Characteristics of paper $i$ for agent $j$

- **Quality and fit**: A paper $i$'s underlying quality is represented by $q_i$, following *Beta*(1,6) in the paper population. Underlying quality does not have an index $j$ because it is assumed to be identical for all agents. Fit denotes the *substantive* usefulness of paper $i$ for agent $j$, represented by $fit_{i,j}$. Fit is expected to vary across agents due to differences in topic or preferences, *i.e.*, agents may choose not to read even a terrific paper if it is on too unrelated a topic. For homogeneous agents, $fit_{i,j} = 0$. $fit_{i,j}$ will raise or lower the quality $q_{i,j}$ of paper $i$ in agent $j$'s eyes as in Equation 1:

$$q_{i,j} = q_i + fit_{i,j} \tag{1}$$

- **Perception error:** represented by $\varepsilon_{i,j}$. It is the error agent $j$ makes in perceiving paper $i$'s quality. While a paper's rhetorical value (see below) and fit are relatively easy to judge from skimming, quality is more difficult to judge and is initially perceived with error. This perception error disappears after an agent reads the paper.
- **Perceived quality:** represented by $s_{i,j,t}$, it denotes the quality of paper $i$ as perceived by agent $j$ at time $t$, before reading. We assume that the higher a paper's citation count $c_{i,t}$ the higher its perceived quality, with the premium determined by a parameter α, i.e., reinforcement strength. Adding all the determinants of a paper's perceived quality yields Equation 2:

$$s_{i,j,t} = q_i + fit_{i,j} + \alpha \times c_{i,t} + \varepsilon_{i,j} \tag{2}$$

- **Overall rhetorical value**: represented by $r_{i,j,t}$, denotes the rhetorical usefulness of paper $i$ for agent $j$ at time $t$. Unlike quality, which has an underlying-quality component $q_i$, rhetorical value is assumed to vary substantially from person to person. For example, a paper taking a side on a debate might be rhetorically useful to those on the same side but not on the other. For heterogeneous agent $j$, paper $i$ has an initial person-specific rhetorical value $r_{i,j}$. For homogenous agents, $r_{i,j} = r_i$. Perceiving the rhetorical value does not require a careful reading so we include a perception error. We assume that rhetorical value increases with perceived quality $s_{i,j,t}$ as in Equation 3:

$$r_{i,j,t} | unread = r_{i,j} + \beta \times s_{i,j,t} \tag{3}$$

The parameter β determines the strength of the reinforcement process of $s_{i,j,t}$. Note, if the agent has not read the paper, her $s_{i,j,t}$ is affected by the perception error in the perceived quality, as shown in Equations 2. If the agent has read the paper closely, the perception error in perceived quality disappears, and the rhetorical value becomes

$$r_{i,j,t} | read = r_{i,j} + \beta \times (s_{i,j,t} - \varepsilon_{i,j}) \tag{4}$$

A crucial question is how quality and rhetorical values are distributed, and parameter values are initialized. We ground the distributions using expert ratings from peer review. Focusing on the prominent *ICML* conference[3], the reviewing platform *OpenReview.net* provides ratings of *submitted* papers. The ratings averaged across a paper's reviewers follow an approximately normal distribution. No papers received maximum points or 0. Assuming that only papers that score in the top 20-30% are accepted (published) leads to a long-tail distribution of ratings of *published* papers[4]. Similarly, team performance in large-scale real-world data exhibits a long-tail distribution rather than a normal distribution (Bradley & Aguinis, 2022). We thus use a beta distribution β(1, *w*) for both quality and initial rhetorical value. Parameter *w* determines the fatness of the distribution's tail, which we set at *w*=6 as it best fits the distributions of ratings of 2019-2022 submissions to *ICML*. In *Appendix Section 1.1,* we test the robustness of our results to different choices of distributions and *w*'s and find that they are qualitatively consistent with the main choices.

## Turning rhetorical citing "on" and "off"

In what we call the "full" model, agents cite substantively *and* rhetorically. At each time period *t* an agent joins the environment. Agents first rank all papers by perceived quality $s_{i,j,t}$. Then, they read the *m* highest perceived-quality papers. After reading the papers, agents observe the papers' actual quality $q_{i,j}$ (with fit) in their eyes (no perception error) and proceed to the citing stage, which occurs in two steps. First, adhering to academic norms, agents substantively draw inspiration from and cite all sufficiently good papers in their eyes where quality $q_{i,j} > \tau_j$. If there are any remaining slots in the citing budget *n*, agents rank all papers on rhetorical value $r_{i,j,t}$ and populate the slots with those with highest $r_{i,j,t}$'s. We allow an agent to select the same paper for a substantive and a rhetorical slot.

We compare this full model to two null models, where agents cite only substantively, *i.e.*, according to $q_{i,j}$. The two null models differ on how they treat the case where there are insufficient papers of high enough quality to fill the entire citing budget.

---

[3] https://openreview.net/group?id=ICML.cc. Accessed 2022-11-01.

[4] Note that because our model depends heavily on papers of the highest value, it is not sensitive to the shape of the distribution for low-quality papers.

- *Null-fixed-reference.* Null model with fixed citing budget: Agents cite the *n* papers with highest $q_{i,j}$, even if they are below the threshold.
- *Null-fixed-threshold.* Null model with fixed threshold**:** Agents cite only papers with $q_{i,j}$ above the threshold, even if that leaves unfilled slots in the reference list.

Figure 1 illustrates the modeling approach.

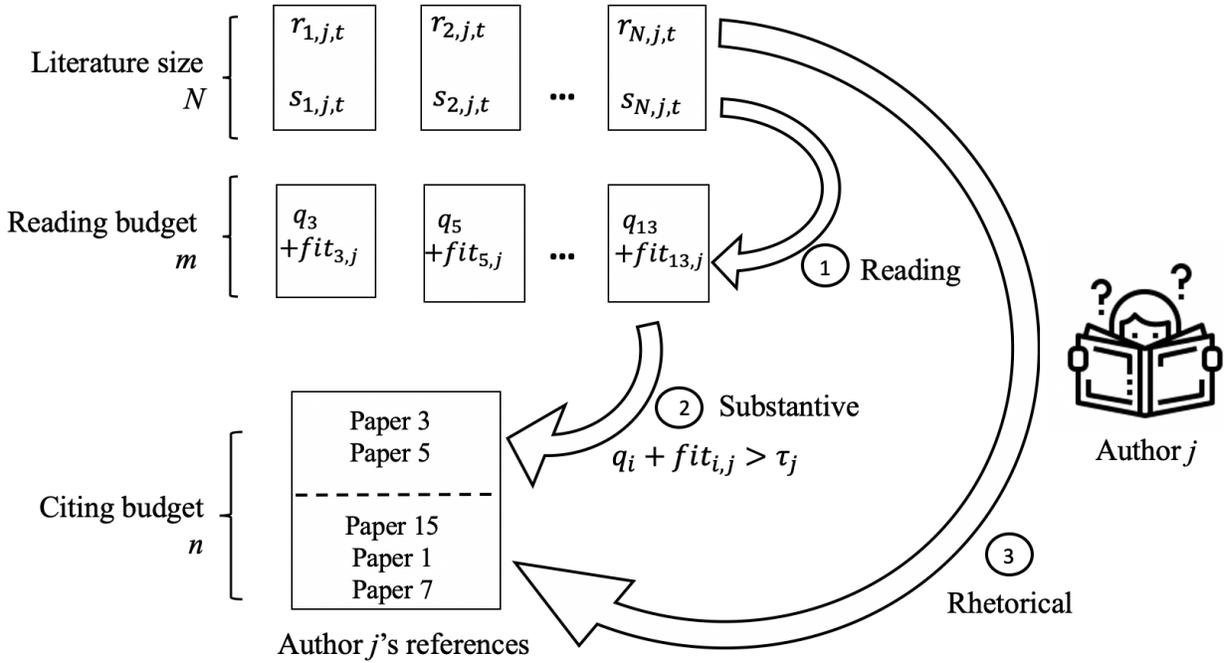

$$C_{i,t+1} = C_{i,t} + 1 \text{ where } i \in \{3, 5, 15, 1, 7\}$$

*Figure 1.* **The citing model**. *Stages are denoted by circled numbers. For each paper i in the literature (N papers), author j observes its perceived quality $s_{i,j,t}$ and rhetorical value $r_{i,j,t}$ at each time step t.* **Stage 1.** *An author chooses the m papers with the highest $s_{i,j,t}$'s to read. Reading reveals the papers' quality in people's eyes $q_i + fit_{i,j}$.* **Stage 2.** *An author chooses all papers with sufficient quality and fit for substantive citations, i.e., $q_i + fit_{i,j} > \tau_j$.* **Stage 3.** *If there are any remaining slots in the reference list, the author rhetorically cites papers with the highest rhetorical values $r_{i,j,t}$ until there are no more slots. Finally, the citation count of each paper is updated and timestep is advanced by 1.*

## Measures

We operationalized the metrics of community health in the following way. To measure the *correlation between citation counts $c_{i,t}$ and quality $q_i$ at time step t* we used the Pearson correlation coefficient. To *measure citation churn at time step t*, we used the number of papers cited in time *t* that were not cited in time *t*-1. Larger values represent more churn. To measure *citation inequality at time step t*, we utilized the Gini coefficient of citation distribution in time *t*, computed by dividing the area between the equal cumulative distribution of citations and the actual cumulative distribution of citations by the area under the curve of the equal cumulative distribution. Larger values represent more inequality.

## Choosing parameter values

The empirical evidence on how much researchers read ("reading budget") or how many "relevant" papers there are for a particular project is limited. Tenopir *et al.* surveyed university scholars in 2012, and they found the mean of monthly article readings was about 20 (Tenopir *et al.*, 2015). In contrast, how much researchers cite ("citing budget") is readily measurable. According to Lancho-Barrantes *et al.*, the average number of references in papers varies between 20 and 50 across fields (Lancho-Barrantes *et al.*, 2010). To initialize the models, we set literature size to 600, reading budget to 120, citing budget to 40. We explore other parameter choices below and in the *Appendix*.

The models are run for 1000 timesteps. At each timestep, one agent makes her citing decisions (*i.e.*, publishes one paper, and thereby cites several from the literature). Then, the citation counts and the quantities that depend on them (perceived quality and rhetorical value) are updated for all papers. For simplicity, the newly "published" paper is not added to the literature. Thus, any paper from the literature can accrue at most 1000 citations by the end of the run. The model is equivalent if 1000 different agents publish one paper each or one agent publishes 1000 different papers.

# 5. Results

We present results from the more realistic heterogeneous agents models, and for completeness report homogeneous agents results in *Appendix, Section 2*.

## Effect of rhetorical citing on three metrics of community health

Figure 2 shows the results from the main model specifications. The top row shows how the three metrics of community health – quality-citations correlation (Panel A), citation churn (Panel B), and citation inequality (Panel C) – evolve across the 1000 timesteps. The middle row extracts one summary statistic for each curve in the top row and compares them across models. Panel D shows that, at the end of 1000 timesteps, quality and citations are more correlated (+0.026) in the full model than in either of the null models. Panel E shows that churn averaged across the 1000 iterations is 2.36 times higher than the null-fixed-reference and 2.17 times higher than the null-fixed-threshold models. Panel F shows that after 1000 iterations, the Gini coefficient, measuring citation inequality, is 30.8% lower than both null models. These results support our main Hypotheses 1, 2, and 3.

To better understand why rhetorical citing affects the community health metrics in this way, we focus on citations to two groups of papers: "high-quality" (underlying quality $q_i$ in the top 40) and "mid-quality" papers ($q_i$ in the top 41-150). These groups account for 25% of our literature (600 papers) but attract approximately 85% of the citations. We then measure how many and what type of citations (substantive and rhetorical) the groups get in the full *vs.* null models. Substantive citing implies that citations should go to the highest-quality papers. Panels H and I show that for the two null models, that is roughly true, with only about 18% of citations going to mid-quality papers. This minority of citations is accrued due to perception errors and variability of $fit_{i,j}$. In contrast, in the full model (Panel G), the fraction of substantive citations going to the highest quality papers is significantly reduced, and mid-quality papers are more cited, particularly rhetorically. Note that in all models low-quality papers receive very few citations of even the rhetorical kind because their overall rhetorical value is likely low (*e.g.*, low quality/citation counts result in low perceived quality), even if their underlying rhetorical value to an agent is high. Overall, rhetorical citing thus raises the relative "visibility" of medium-quality papers and results in citations no longer being concentrated on a small, stable set of high-quality papers.

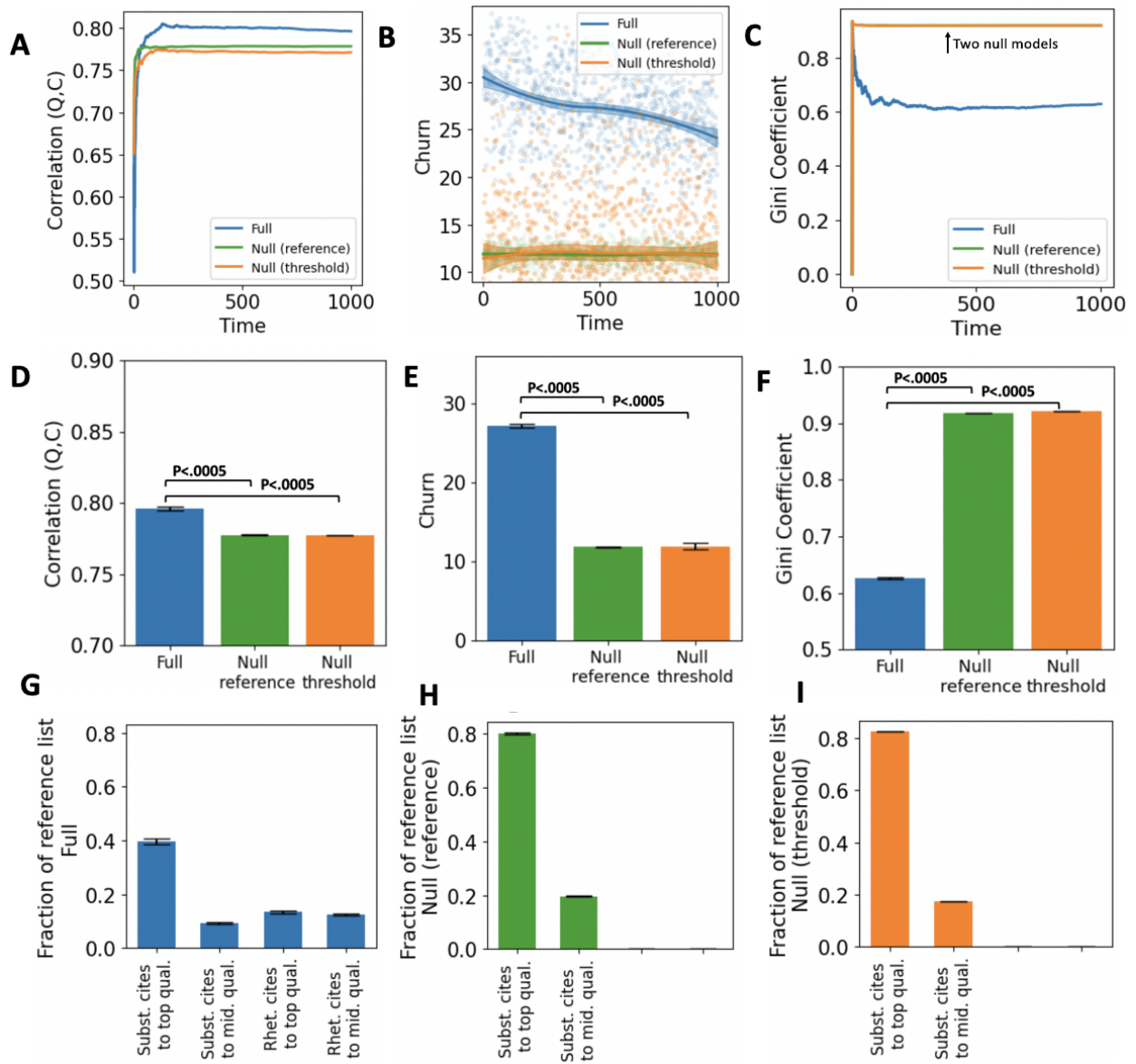

*Figure 2. **A-C:** show how the three metrics of community health (correlation between quality Q and citation count C, citation churn, and citation inequality) evolve across the 1000 iterations of the full, null-reference, and null-threshold models. **D:** compares the quality-citation correlation after 1000 iterations. **E:** compares citation churn averaged over 1000 iterations. **F:** compares citation inequality after 1000 iterations. **G-I:** show the fraction of the reference list (averaged over 1000 iterations) taken up by substantive citations to papers in the top 40 of quality, substantive citations to papers in the top 41-150 of quality, rhetorical citations to the top*

40-quality papers, and rhetorical citations to top 41-150-quality papers. Note: null models (Panels **H, I**) only have substantive citations. Shaded regions represent bootstrapped 95% confidence intervals, and error bars represent conventional 95% confidence intervals.

## Moderating effects of citing budget, reading budget, and literature size

### Citing budget

To understand the role of the citing budget, we fix the reading budget at 120 papers, the literature size at 600 papers, and vary the citing budget from 20 to 100. Figure 3 shows how the three metrics of community health change. In Panels A and C, at each citing budget the model is run ten times for 1000 timesteps each, and the metric is calculated at the end of each run. For Panel B, the metric is averaged over 1000 timesteps in each run.

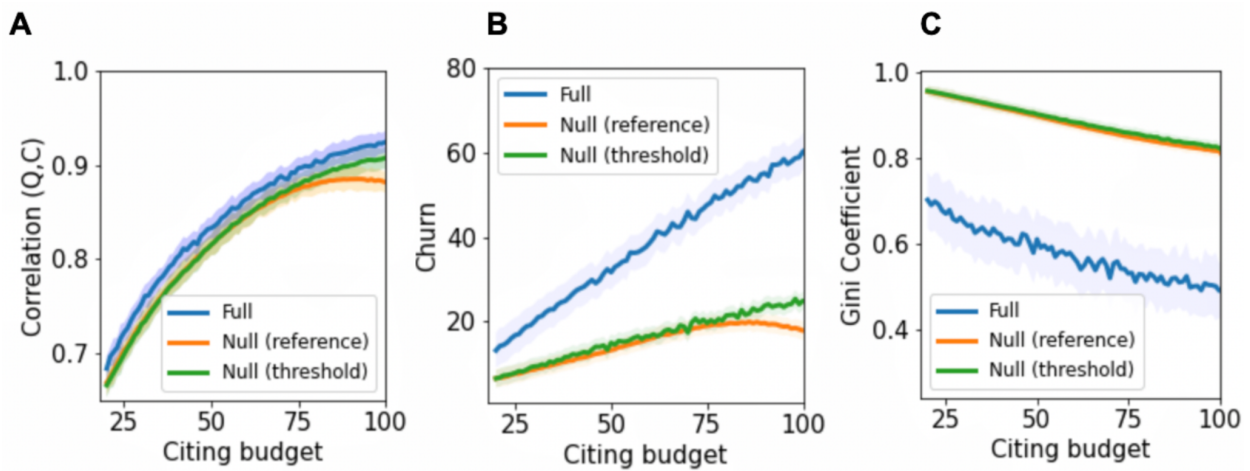

***Figure 3. Citing budget.*** *How increasing the citing budget from 20→100 affects the correlation between citations and quality (Panel A), churn (Panel B), and citation inequality (Panel C). Shaded regions represent bootstrapped 95% confidence intervals.*

Increasing the citing budget from 20 to 100 substantially affects all metrics. In Panel A, the citation-quality correlation increases by 35.3% (0.68 to 0.92, $p < 0.0005$) in the full model, 31.3% (0.67 to 0.88, $p < 0.0005$) in the null-fixed-reference model, and 35.8% (0.67 to 0.91, $p < 0.0005$) in the null-fixed-threshold model. In Panel B, churn increases by 4.59 times (13.12 to 60.23, $p < 0.0005$) in the full model, 2.76 times (6.46 to 17.80, $p < 0.0005$) in the null-fixed-reference model, and 3.83 times (6.47 to 24.81, $p < 0.0005$) in the null-fixed-threshold model. Churn in the full model increases from 2.03 times to 3.38 times higher than the null-fixed-reference model and from 2.03 times to 2.43 times higher than the null-fixed-threshold model. In Panel C, citation inequality decreases by 30.0% (0.70 to 0.49, $p < 0.0005$) in the full model and 14.6% (0.96 to 0.82, $p < 0.0005$) in either null model. These results support Hypothesis 5A. Intuitively, increasing the citing budget gives authors more opportunities to cite less elite-quality but still good-quality papers.

Reading budget

To understand the role of the reading budget, we fix the citing budget at 40, the literature size at 600, and vary the reading budget from 50 to 150. Figure 4 shows how the three metrics of community health change across reading budgets.

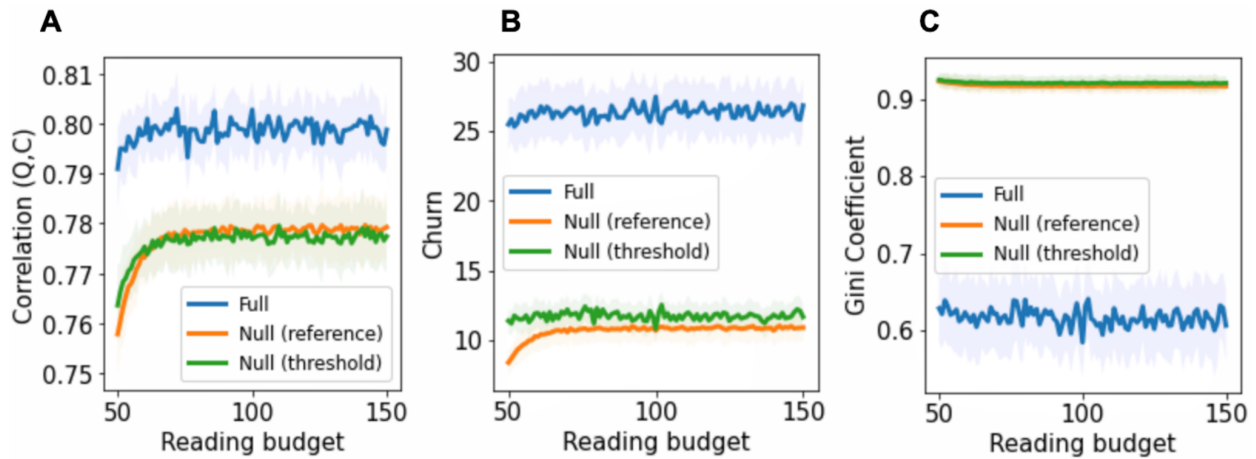

*Figure 4. Reading budget.* *How increasing the reading budget from 50→150 affects the correlation between citation and quality (Panel A), churn (Panel B), and citation inequality (Panel C). Shaded regions represent bootstrapped 95% confidence intervals.*

Unlike citing budget, increasing the reading budget has weaker and more mixed effects on the metrics. The citation-quality correlation increased very weakly in all models (Panel A: full model: 0.791 to 0.798, $p = 0.01$, coefficient =1.723e-05; null-fixed-reference model: 0.757 to 0.779, $p < 0.001$, coefficient = 8.232e-05; null-fixed-threshold model: 0.764 to 0.777, $p < 0.001$, coefficient=4.704e-05). Citation churn changed slightly in three models (Panel B: full model: 25.5 to 26.9, $p = 0.005$, coefficient = 0.0041; null-fixed-reference model: 8.38 to 10.89, $p < 0.001$, coefficient=0.0097; null-fixed-threshold mode: 11.38 to 11.67, $p = 0.938$, coefficient =

-7.512e-05). Citation inequality did not change substantially (Panel C: full model: 0.63 to 0.61, $p$ = 0.016, coefficient=-8.85e-05; null-fixed-reference model: 0.923 to 0.917, $p$ <0.001, coefficient=-2.666e-05; null-fixed-threshold model: 0.926 to 0.922, $p$ < 0.001, coefficient=-1.55e-05). Overall, the reading budget did not have a substantial effect on the three metrics of community health. While it may be epistemically valuable to read more papers, the key constraint on whether those papers get formal recognition in the form of citations is the citing budget. These results support Hypothesis 5B.

Literature size

To understand the role of literature size, we fix the reading budget at 120, the citing budget at 40, and vary the literature size from 200 to 800. Figure 5 shows how the three metrics of community health change across literature sizes.

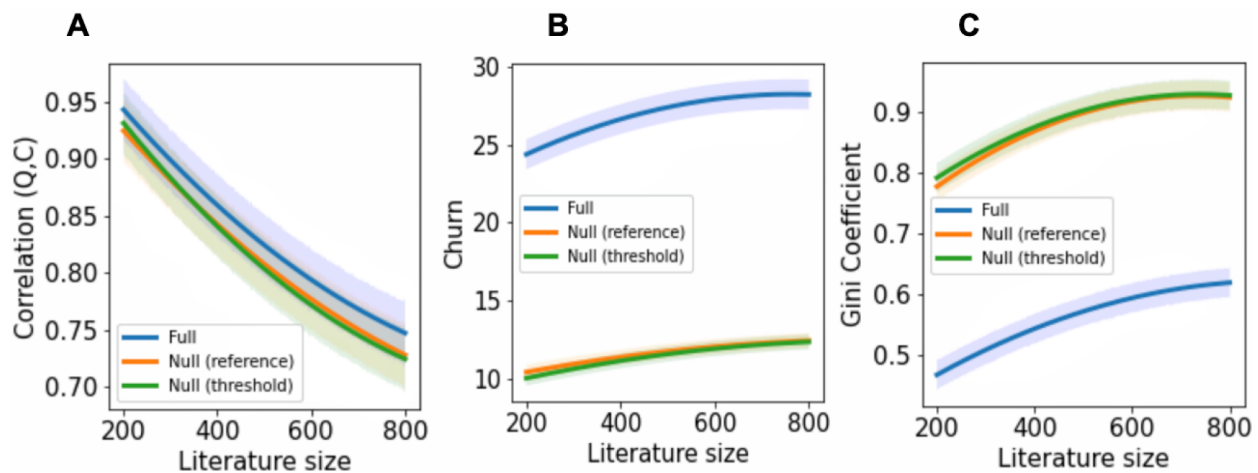

*Figure 5. Literature size.* *How increasing the literature size from 200→800 affects the correlation between citation and quality (Panel A), churn (Panel B), and citation inequality (Panel C). Shaded regions represent bootstrapped 95% confidence intervals.*

Increasing literature size has substantial but mixed effects on the metrics. When the literature size increases from 200 to 800, the citation-quality correlation decreases by 22.1% in the full model (0.95 to 0.74, $p < 0.0005$), 21.7% in the null-fixed-reference model (0.92 to 0.72, $p < 0.0005$), and 22.6% in the null-fixed-threshold model (0.93 to 0.72, $p < 0.0005$). Citation churn increases by 26.3% in the full model (22.52 to 28.43, $p < 0.0005$), 29.3% in the null-fixed-reference model (9.21 to 11.91, $p < 0.0005$), and 22.8% in the null-fixed-threshold model (10.16 to 12.48, $p < 0.0005$). Citation inequality increases by 30.6% in the full model (0.49 to 0.64, $p < 0.0005$), 22.1% in the null-fixed-reference model (0.77 to 0.94, $p < 0.0005$), and 20.5% in the null-fixed threshold model (0.78 to 0.94, $p < 0.0005$). These results support Hypothesis 4.

Overall, increasing literature size decreases the correlation between citations and quality, and increases citation churn and inequality. Intuitively, as literature size grows, the *fraction* of papers read and cited decreases, with both being concentrated on elite-quality pieces. Surprisingly, citation churn increases as well. As the literature size grows, the more high-quality papers (above the threshold) there are. Idiosyncratic factors like $fit_{i,j}$ can make authors substitute one high-quality paper for another.

# 6. Discussion

This paper used agent-based models to understand the effect of rhetorical citing on scientific communities. We simulated worlds with substantive and rhetorical citing, *i.e.,* the current world, *vs.* only substantive citing practices, and measured how turning rhetorical citing "on" and "off" affected three relatively measurable and much-discussed metrics of community health – the correlation between citations and quality, citation churn, and citation inequality. Surprisingly, we found that rhetorical citing increases the degree to which citation counts correlate with a paper's quality, increases citation churn, and decreases citation inequality. The proximate explanation for these effects is that rhetorical citing redistributes citations from the few elite-quality papers to a more diverse set. The more fundamental explanation is that when seeking papers to cite rhetorically, researchers select on factors beyond just quality (which may still be important), such as rhetorical value.

Furthermore, increasing the length of reference lists ("citing budgets"), usually seen as a problem, increased churn, citation-quality correlation, and decreased citation inequality. Increasing the reading budget, usually encouraged, had little effect on three metrics. Lastly, increasing the literature size had mixed effects, increasing churn, decreasing citation-quality correlation, and increasing inequality. These additional results further show the value of modeling how intuitive recommendations from the scientific community may have underappreciated consequences.

Models such as ours necessitate many simplifications and scope conditions, which we believe are fruitful directions for future research. First and foremost, the three metrics of community health we focus on are not exhaustive or unambiguous in interpretation. While it is relatively unambiguous that a higher citation-quality correlation is better for science than a lower one, the optimal levels of churn and inequality are more ambiguous, although the current literature raises concerns that the current levels are higher than optimal (Chu and Evans 2021; Nielsen and Andersen 2021). If one takes the conservative view that only one of our three metrics has an unambiguous interpretation, then a measured conclusion of our results is that rhetorical citing has some effects that are plausibly positive. However, other potential metrics, like the amount of misinformation in the literature or misallocation of rewards, were not included (West & Bergstrom, 2021). We hope our work stimulates effort to model more dimensions of community health to better capture the overall effects of a different world.

Second, we assumed that the agents and the types of papers they produce remain fixed across the different worlds. In other words, we did not account for how agents might change their behavior in response to the types of citation practices that exist in a community. For example, in a world with only substantive citing, agents may seek to produce papers of high quality rather than high rhetorical value. Note that while incentives that induce papers of high quality only may appear to be ideal, such a world may have the same concerns as the substantive-citing worlds this paper explored. Third, as noted previously, our models are designed to capture reading and citing dynamics in the short-to-medium term. In the long-term, it is important to take into account the scientific obsolescence of older works, the decline in the reinforcing strength of previous citations, the addition of newly published papers, and other signals of quality, *e.g.*, journal impact factor.

# 7. Conclusion

Citing papers for reasons other than to acknowledge their influence ("rhetorical citing") is widely discouraged because it is assumed to harm scientific communities. The assumption is intuitive and supported by numerous examples of harmful effects (Greenberg, 2009; Letrud & Hernes, 2019; Rekdal, 2014; De Vries *et al.*, 2018). It is thus very tempting to envision the counterfactual world with only substantive citing practices optimistically. The first contribution of the paper is to model that counterfactual world realistically, and compare it to the present. The comparison revealed that rhetorical citing changes, and arguably improves, three aspects of scientific community health. By redistributing citations from a stable set of elite-quality papers to a broader set, rhetorical citing increases the extent to which citations measure quality and increases dynamism in scientific literatures. While previous work has pointed to the volume of research as a driver of stagnation (Chu & Evans, 2021), our work reveals that some seemingly "bad" practices in science, like citing papers without being influenced by them, can help mitigate it. This finding points to a broader conclusion – citations are the outcome of a longer process, driven by how researchers search for and read papers. Consequently, attempts to improve only the last part of this citing process without improving the earlier steps may be of limited utility and may even have unintended consequences.

The study did not aspire to measure the *total* effect of rhetorical citing. Such an analysis would require taking into account many more direct and indirect effects, such as misinformation in the literature (Cobb *et al*., 2023; West & Bergstrom, 2021). Thus, the paper does not endorse rhetorical citing, but only argues that it is not a *priori obvious* whether the scientific community is better off without it. More broadly, the study suggests that when considering policies to fix a particular dysfunction in research, it is important to account for the broad set of incentives and cognitive constraints within which researchers operate.

Performing the comparison above necessitated the development of a behavioral model of citing, in which researchers are cognitively constrained and cite for both substantive and rhetorical reasons. Such a synthetic model of citing has proven to be an elusive goal for decades of scientometrics and information science literature (Bornmann & Daniel, 2008; Cronin, 1981), with limited progress to this day. The paper's second contribution is to offer one such synthetic model, drawing on the rich empirical literature on how researchers search for, read, and cite papers.

**Acknowledgements**: We are grateful to Charles Gomez, Charles Ayoubi, Inna Smirnova, Wei Yang Tham, and seminar participants at the Digital, Data, and Design Institute at Harvard Business School for helpful feedback. All errors are our own.

# Appendix for Do "bad" citations have "good" effects?

Honglin Bao, Misha Teplitskiy

## Table of contents

# 1. Robustness checks

The Main paper considered several moderation variables, *e.g.*, literature size. Here, as we vary various parameters to test for robustness, we also parameter-sweep arguably the most important and policy-relevant moderator – the citing budget. As we change the focal parameters, we show how the results vary as the citing budget goes from 20 to 100 and includes more rhetorical citations.

## 1.1. Varying quality and initial rhetorical value distributions

In the Main model, we assumed long-tail distributions for quality and initial rhetorical value. The tail's fatness may have a significant effect on three metrics of community health. Here, we examine the robustness of our principal conclusions using different value distributions. In the Main model we used β(1, *w*) for distributions of quality and rhetorical value. Parameter *w* determines the fatness of the distribution's tail, which we set *w* at 6. We vary it to 4 and 8 to see how the results will change with more/fewer high-value papers, respectively. In addition, we consider a normal distribution *N*(0.5, 0.1). All distributions are shown in Figure S1.

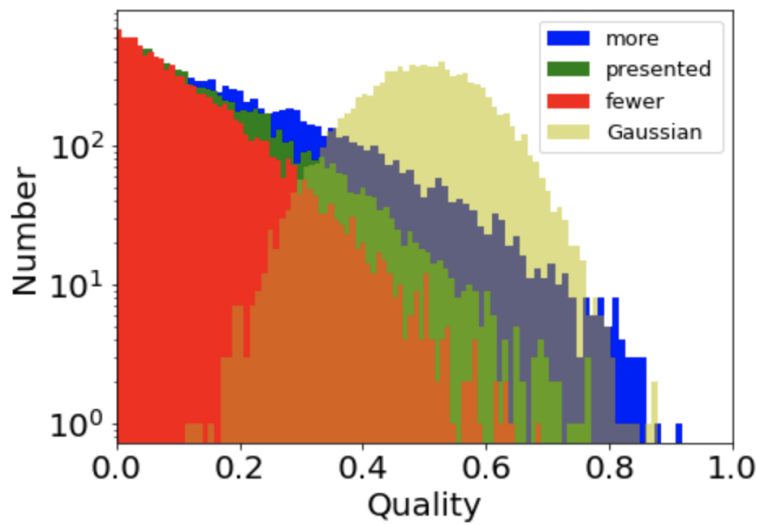

*Figure S1. Value distributions.* *We tested three additional distributions for quality and initial rhetorical value: more high-value papers β(1, 4), fewer high-value papers β(1, 8), and values following a Gaussian distribution N(0.5, 0.1) among papers. The presented model in the Main paper uses a Beta distribution β(1, 6) of values with moderate numbers of high-value ones.*

## More high-value papers

Using a distribution with more papers of high quality and rhetorical value ($w = 4$), we find that rhetorical citing still improves the three metrics of health, consistent with the Main results. Figure S2 plots how the metrics for the full and two null models change across different citing budgets. The result for each citing budget is the measurement after a 1000-time-step simulation, averaged by ten random seeds.

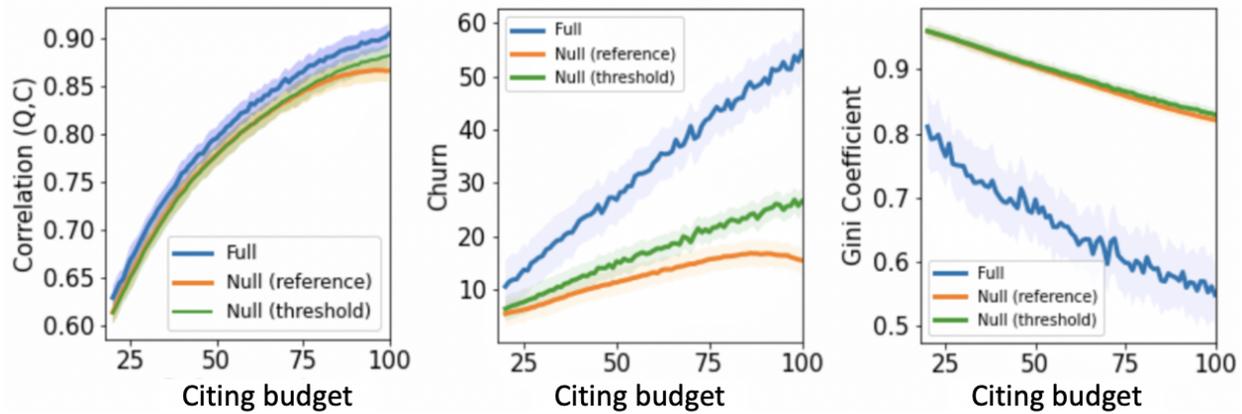

*Figure S2.* *The results with more high-value papers β(1, 4). Shaded regions represent bootstrapped 95% confidence intervals.*

## Fewer high-value papers

Similarly to the above, using a distribution with fewer papers of high quality and rhetorical value ($w = 8$), we find that rhetorical citing still improves the three metrics of health, consistent with the Main results.

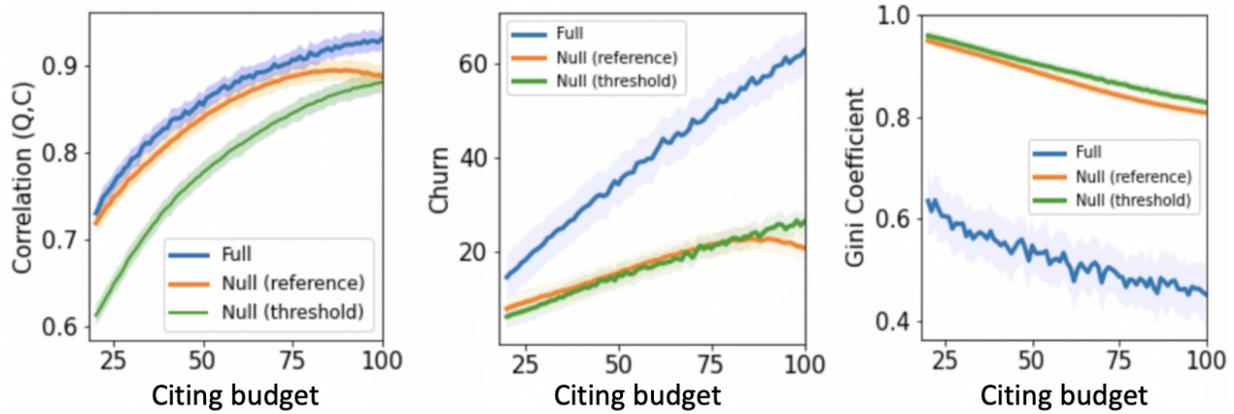

*Figure S3*. *The results with fewer high-value papers β(1, 8). Shaded regions represent bootstrapped 95% confidence intervals.*

Normal distribution of values

Next, we consider the case where quality and rhetorical value are distributed normally. The full model (with rhetorical citing) improves the metrics relative to the null models, especially for higher citing budgets.

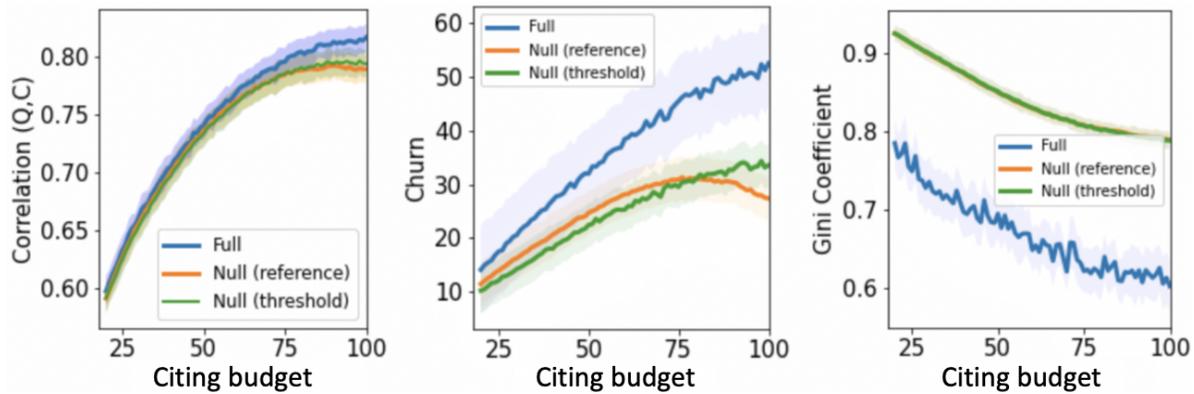

*Figure S4.* *The results under a normal distribution of values N(0.5, 0.1). Shaded regions represent bootstrapped 95% confidence intervals.*

## 1.2. Varying distributions of threshold

In the Main results, we sampled adoption thresholds for agents from a uniform distribution on [0,1]. Here, we instead use a truncated normal distribution *N*(0.5, 0.2) within the range [0,1].

Figure S5 shows that rhetorical citing improves the three metrics, consistent with the Main results.

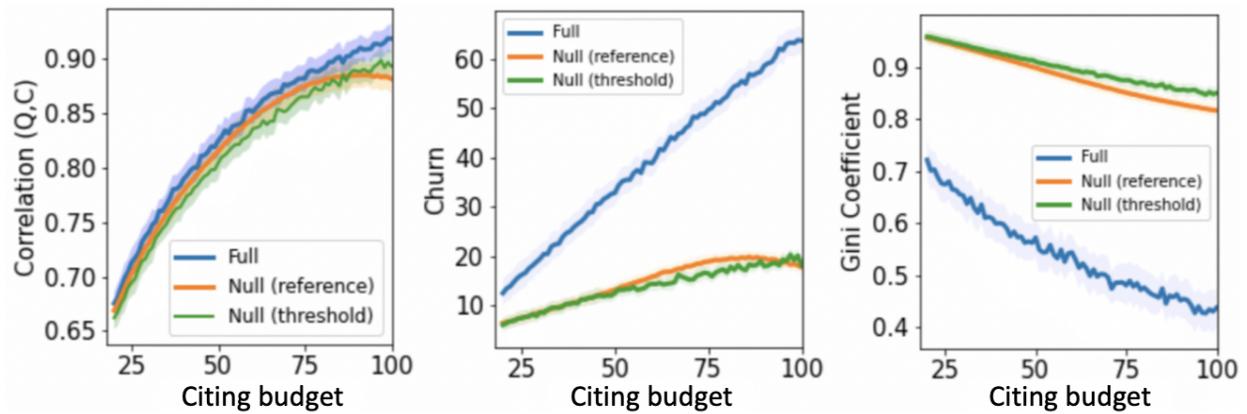

*Figure S5.* *Three models with different thresholds, N(0.5, 0.2) truncated within [0,1]. Shaded regions represent bootstrapped 95% confidence intervals.*

## 1.3. Varying size of perception error

In the Main result, perception error was distributed *Normal*(0, 0.05). Here, we consider how the results change when the standard deviation is lowered to 0.02 or raised to 0.1.

Higher noise

Figure S6 shows that rhetorical citing improves the three metrics, consistent with the Main results.

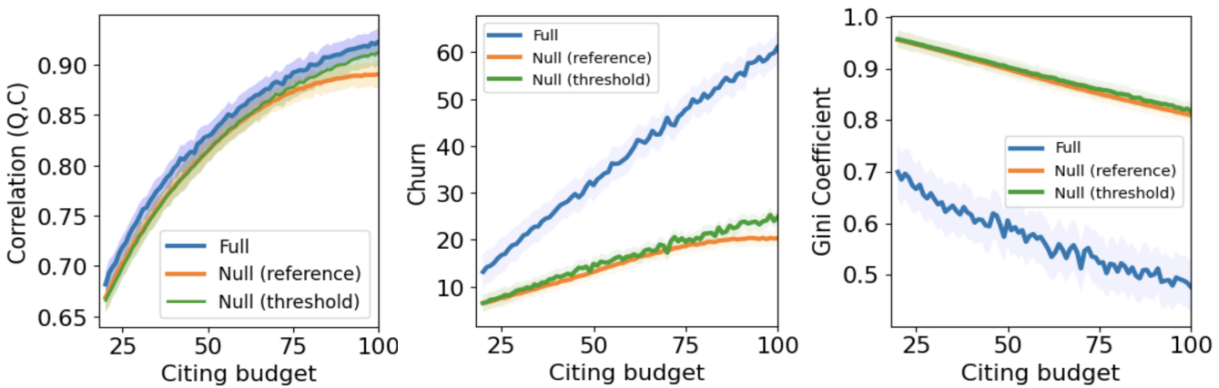

*Figure S6.* *Three metrics of community health with higher perception noise, Normal(0, 0.1). Fit and perception errors can vary a paper's quality from 0.5 to a random number in [0.1, 0.9]. Shaded regions represent bootstrapped 95% confidence intervals.*

Lower noise

Figure S7 shows that rhetorical citing increases the correlation (left panel), churn (middle panel), and decreases citation inequality (right panel), consistent with the Main results.

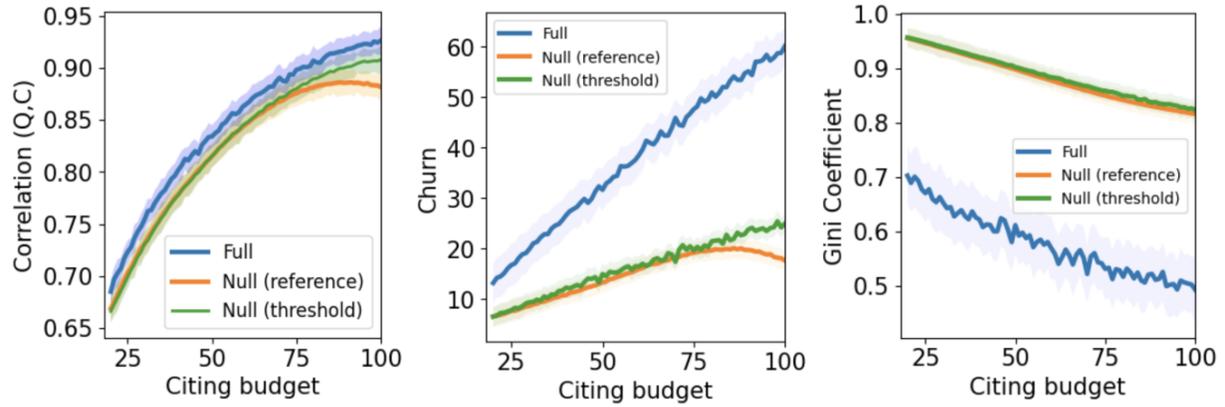

*Figure S7.* Three metrics of community health with lower perception noise, Normal(0, 0.02). Fit and perception errors can vary a paper's quality from 0.5 to a random number in appx. [0.35, 0.65]. Shaded regions represent bootstrapped 95% confidence intervals.

## 1.4 Varying distribution of *fit*

In the Main result, we sampled *fit* for agents from a uniform distribution on [-0.1,0.1]. Here, we consider how the results change when the range of the distribution is increased to [-0.2,0.2] or decreased to [-0.05,0.05].

Less varied *fit*

Figure S8 shows that rhetorical citing improves the three metrics, consistent with the Main results.

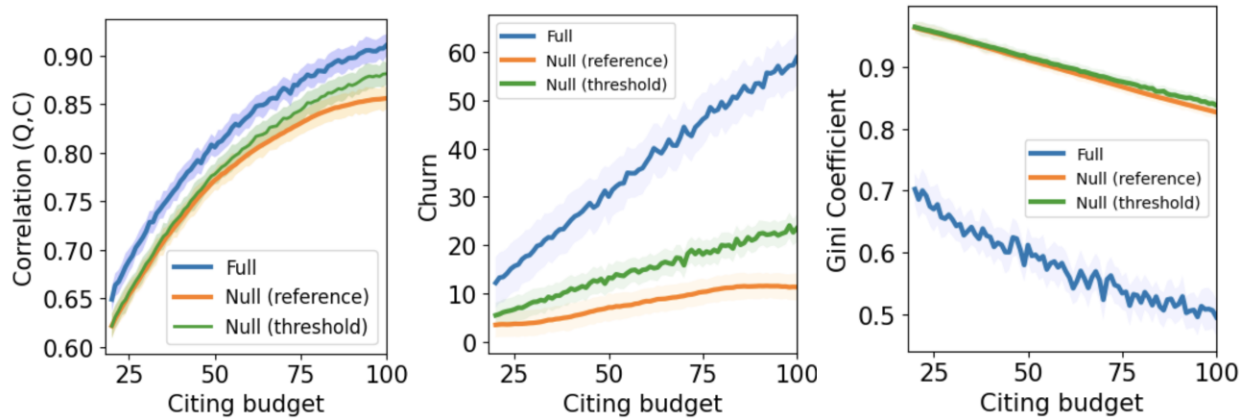

*Figure S8.* Three metrics of community health with a less varied fit, Uniform(-0.05, 0.05). Fit and perception errors can vary a paper's quality from 0.5 to a random number in [0.3, 0.7]. Shaded regions represent bootstrapped 95% confidence intervals.

More varied *fit*

Figure S9 shows that rhetorical citing improves churn (the middle panel) and citation inequality (the right panel). However, the citation-quality correlation changes little between the full and null models. The intuition behind this result is that with a more varied *fit*, existing perceived quality of papers does not much affect whether researchers cite them subsequently as varied fits can significantly raise or lower the quality in readers' eyes. Hence, the lock-in effect observed in the null models in our Main results is reduced. In effect, rhetorical citing has similar but much stronger effects on the community health metrics as adding substantial person-specific purposes to the citing process.

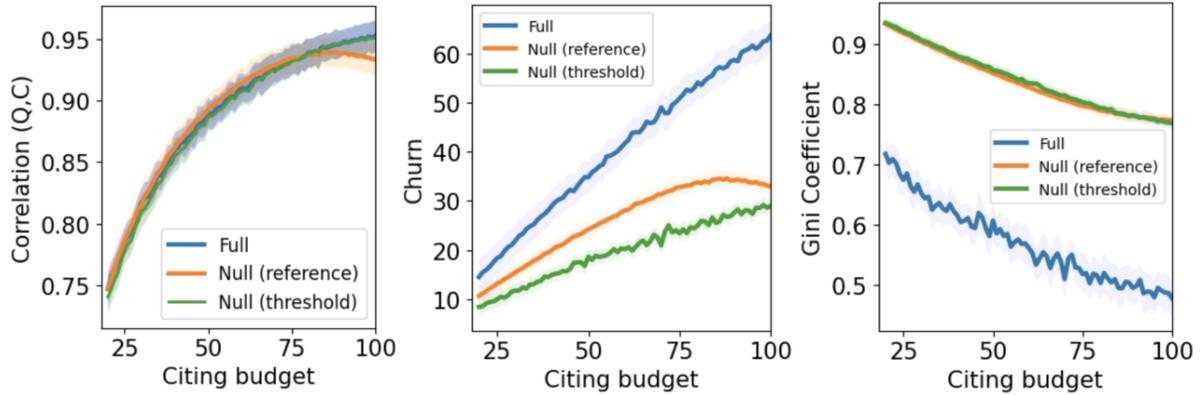

*Figure S9.* Three metrics of community health with a more varied fit, Uniform(-0.2, 0.2). Fit and perception errors can vary a paper's quality from 0.5 to a random number in [0.15, 0.85]. Shaded regions represent bootstrapped 95% confidence intervals.

## 1.5 Varying reinforcement strengths

Varing α

The parameter α measures how a paper's citation count affects perceptions of its quality, with $\alpha = 0.001$. Note that because the maximum citation count is 1000, the maximum effect of citations on perceived quality is 0.001*1000 = 1, which equals the maximum underlying quality and maximum underlying rhetorical value. Here, we consider the case $\alpha = 0.002$ (maximum citation premium=2) and $\alpha = 0$ (citations have no impact on perceived quality).

Figure S10 shows that the full model outperforms the two null models in the three metrics if readers do not rely on citations to perceive quality. In contrast, Figure S11 shows that under a higher reinforcement strength, the full model outperforms the two null models on churn and inequality, but is similar on correlation. Greater reinforcement increases the concentration of substantive citations among elite-quality papers, and because perceived quality is a component of rhetorical value, increases rhetorical citations on them as well, diminishing the differences between the models.

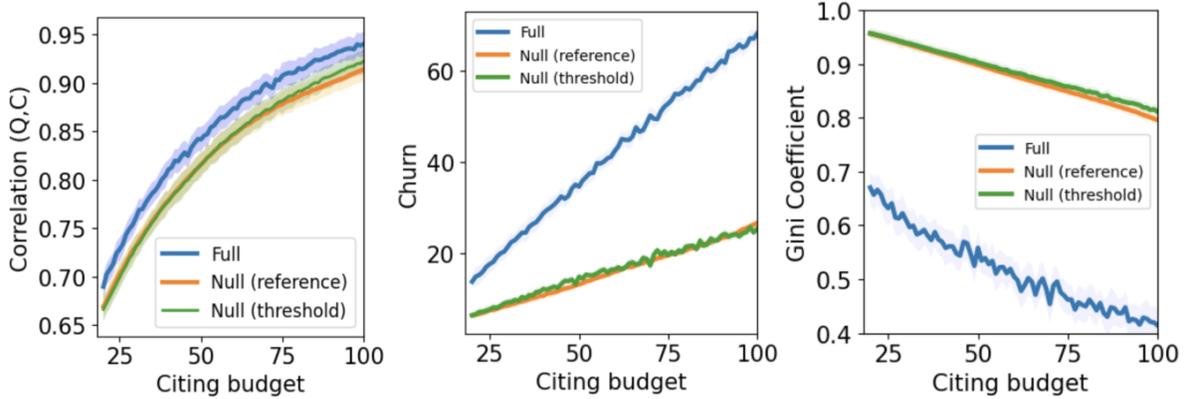

***Figure S10.*** *Three metrics under zero reinforcing strength (α = 0). Shaded regions represent bootstrapped 95% confidence intervals.*

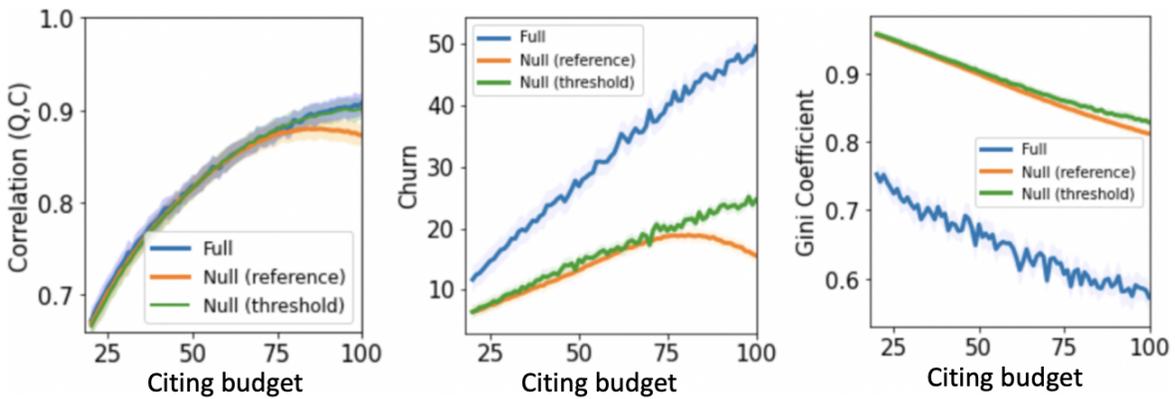

***Figure S11.*** *Three metrics under a higher reinforcing strength (α = 0.002). Shaded regions represent bootstrapped 95% confidence intervals.*

Varing β

The parameter β quantifies how a paper's perceived quality affects its rhetorical value, with $β = 0.3$ in the Main model. Here we compare the Main model with β = 0, where rhetorical value does not depend on perceived quality at all, and β = 1, where in determining rhetorical value agents place equal weight on how rhetorically useful the paper is to their argument and its perceived quality. Note that because the null models do not have rhetorical citing, we do not include them in this analysis.

Figure S11 shows that the full model with β = 0 has the highest citation churn and the lowest citation inequality. The full model with β = 1 has the lowest churn and highest inequality. Interestingly, the model with the intermediate β = 0.3 value that we use in Main has the highest citations-quality correlation. If agents rely on perceived quality very heavily, then the model comes close to the null models without rhetorical citing, lowering the correlation as is observed in the null models. On the other hand, if agents do not rely on perceived quality at all, then the underlying rhetorical value becomes the only driving factor in citing and even low-quality papers receive citations. Thus, very high and very low reinforcement induce lower citations-quality correlations by increasing citing of the upper and lower ends of the quality distribution, respectively.

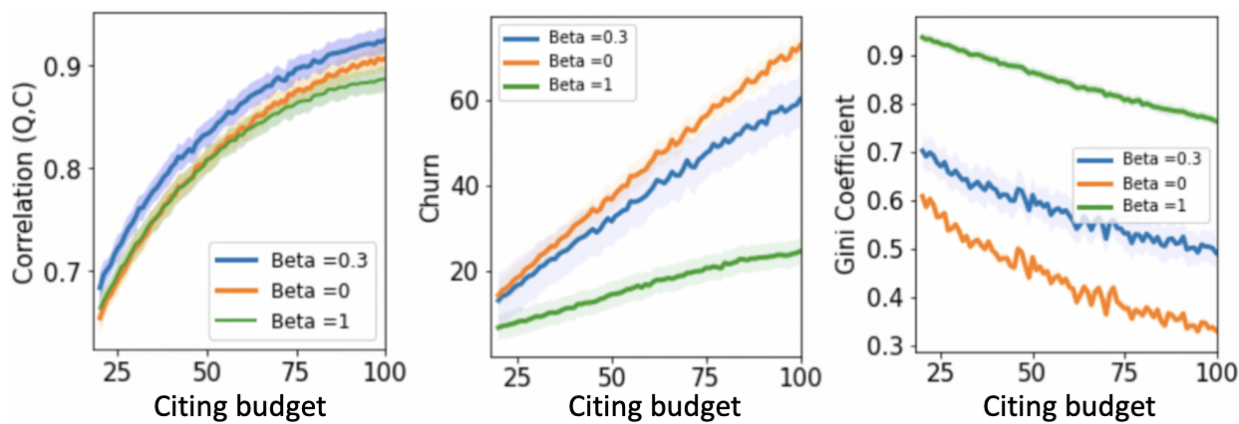

*Figure S11.* *Full models after 1000 iterations with different values of β. Shaded regions represent bootstrapped 95% confidence intervals.*

## 2. Model with homogeneous agents

In contrast to the heterogeneous agent models presented in Main, homogeneous agents perceive quality identically (there is no $fit_{i,j}$), perceive rhetorical value identically, and have the same threshold for adoption 0.5. We initialize models with homogeneous agents the same way as with heterogeneous ones: literature size = 600, reading budget = 120, citing budget = 40, and timesteps = 1000. Figure S12 shows the citation distribution across paper quality. After 1000 iterations, the distribution is nearly bimodal with some papers having 0 citations and some 1000. Given that citation distributions in practice are never bimodal, we do not believe such a simple model is sufficiently realistic. For completeness, we present all three community health metrics for the three versions of this model (Full, Null-reference, Null-threshold) in Figure S13, but do not interpret it further.

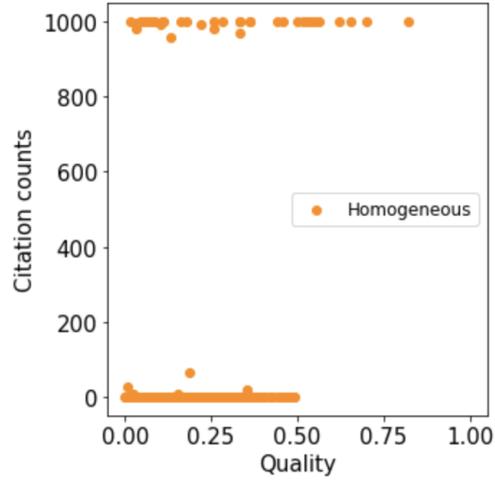

*Figure S12.* Citation distribution in the homogeneous model after 1000 iterations.

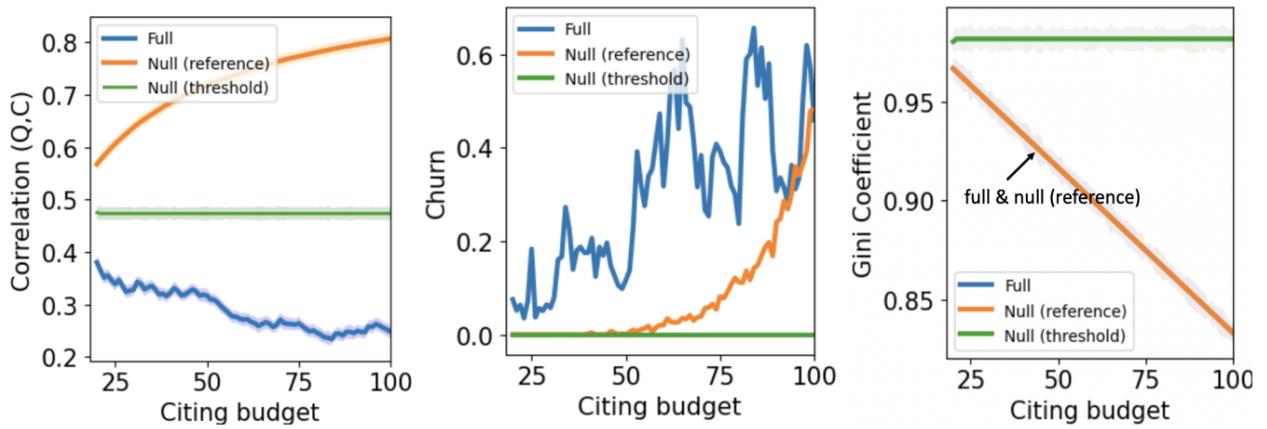

*Figure S13.* Metrics of community health produced by three models – Full, Null-reference, Null-threshold – all with homogeneous agents, after 1000 iterations. Shaded regions represent bootstrapped 95% confidence intervals.